\documentclass[conference]{IEEEtran}
\IEEEoverridecommandlockouts
\usepackage{cite}
\usepackage{amsmath,amssymb,amsfonts}
\usepackage{graphicx}
\usepackage{textcomp}
\usepackage{xcolor}

\usepackage{algpseudocode}
\usepackage{tabu}
\usepackage{multirow}
\usepackage{amsthm}

\DeclareGraphicsExtensions{.pdf,.eps}
\usepackage{subcaption}

\usepackage{listings}
\usepackage{url}
\def\BibTeX{{\rm B\kern-.05em{\sc i\kern-.025em b}\kern-.08em
    T\kern-.1667em\lower.7ex\hbox{E}\kern-.125emX}}
\begin{document}

\title{Trash Talk: Accelerating Garbage Collection on Integrated GPUs is Worthless\\
}

\author{\IEEEauthorblockN{Mohammad Dashti}
\IEEEauthorblockA{\textit{University of British Columbia}\\
Vancouver, Canada \\
mdashti@ece.ubc.ca}
\and
\IEEEauthorblockN{Alexandra Fedorova}
\IEEEauthorblockA{\textit{University of British Columbia}\\
Vancouver, Canada \\
sasha@ece.ubc.ca}
}

\maketitle

\begin{abstract}
Systems integrating heterogeneous processors with unified memory provide seamless integration among these processors with minimal development complexity. These systems integrate accelerators such as GPUs on the same die with CPU cores to accommodate running parallel applications with varying levels of parallelism. Such integration is becoming very common on modern chip architectures, and it places a burden (or opportunity) on application and system programmers to utilize the full potential of such integrated chips. In this paper we evaluate whether we can obtain any performance benefits from running garbage collection on integrated GPU systems, and discuss how difficult it would be to realize these gains for the programmer. 

Proliferation of garbage-collected languages running on a variety of platforms from handheld mobile devices to data centers makes garbage collection an interesting target to examine on such platforms and can offer valuable lessons for other applications. We present our analysis of running garbage collection on integrated systems and find that the current state of these systems does not provide an advantage for accelerating such a task. We build a framework that allows us to offload garbage collection tasks on integrated GPU systems from within the JVM. We identify  dominant phases of garbage collection and study the viability of offloading them to the integrated GPU. We show that performance advantages are limited, partly because an integrated GPU has limited advantage in memory bandwidth over the CPU, and partly because of costly atomic operations.
\end{abstract}

\begin{IEEEkeywords}
GPGPU, Garbage collection, Heterogeneous systems
\end{IEEEkeywords}

\section{\textbf{Introduction}}

\begin{figure}[!ht]
\centering
\includegraphics[width=\linewidth]{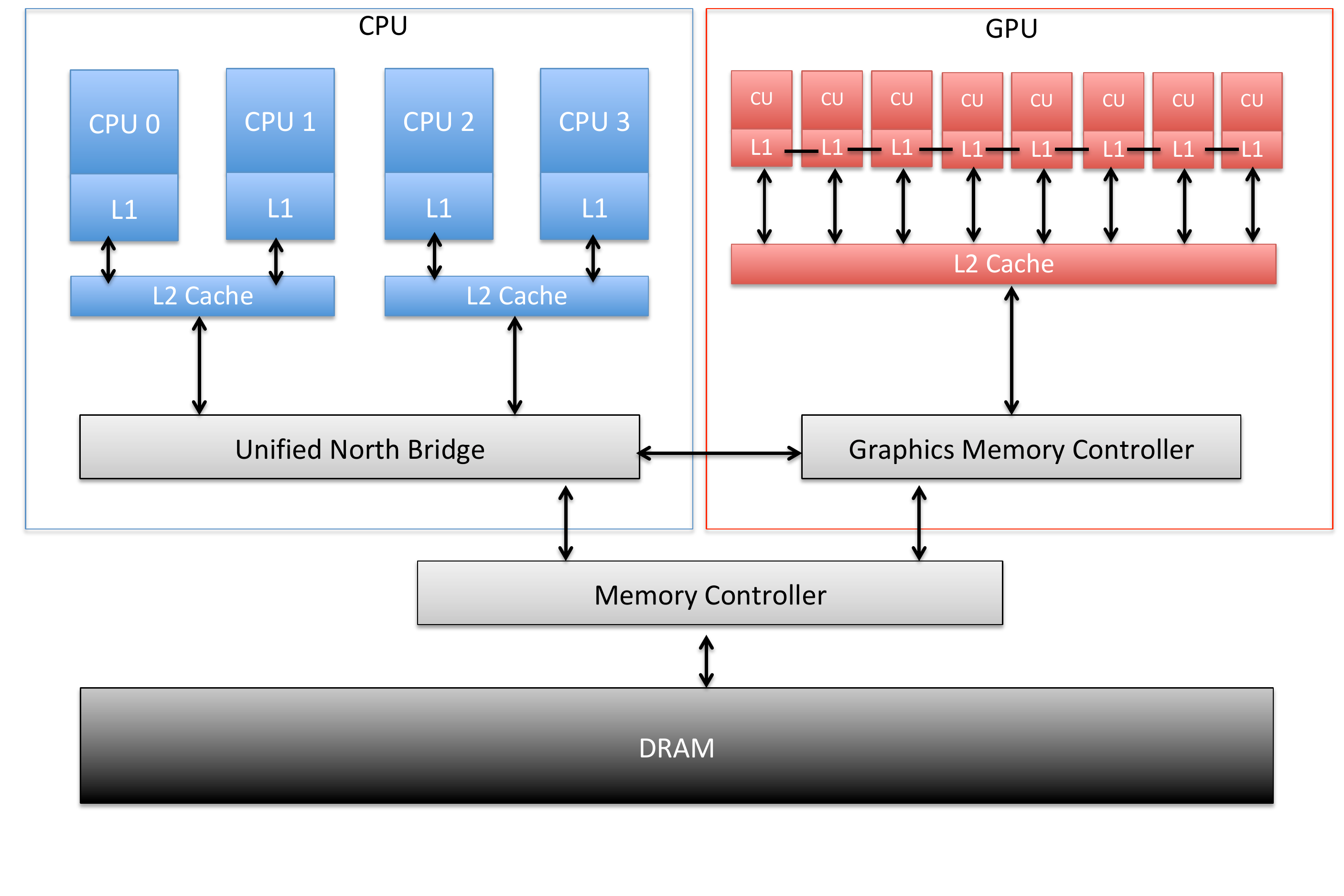}
\caption{Abstract memory system architecture of the integrated system used in our work.}
\label{fig:kaveri}
\end{figure}
High performance and energy efficiency are the ultimate goals of any computing system. However, as Dennard’s scaling~\cite{dennard1974design,esmaeilzadeh2011dark} comes to an end, chip designers can no longer rely on increasing transistor density and clock frequencies to boost performance within a fixed power envelope. The focus has turned to multicore chip designs where multiple symmetric or asymmetric cores occupy the chip area~\cite{dashti2017analyzing,navarro2019heterogeneous,agostini2020balancing,rodriguez2019parallel}. One of the emerging hardware architectures in the asymmetric realm places the GPU and the CPU \emph{on the same die}, sharing the same physical memory (figure~\ref{fig:kaveri}). This type of architecture can eliminate data transfer costs compared to traditional discrete systems where the accelerator (GPU) sits on a separate chip connected to the CPU through a Peripheral Component Interconnect (PCI) bus. Such tight integration enables the CPU and the accelerator to share and collaborate on data without explicit copying. The elimination of data copies can significantly improve performance even compared to much more powerful discrete GPUs. For compute-intensive applications, a discrete GPU is generally much better due to its superior compute capabilities. However, for applications that need to transfer large amounts of data, a less powerful and more energy-efficient integrated GPU can significantly outperform a discrete GPU~\cite{dashti2017analyzing,spafford2012tradeoffs,daga2012exploiting}.

In this paper, we aim to answer the question of whether we can obtain  performance benefits from offloading garbage collection (GC) in the JVM onto a GPU in integrated CPU/GPU systems. We approach this question via a case study of four memory intensive applications that are affected by GC: Apache Spark~\cite{zaharia2016apache}, GraphChi~\cite{kyrola2012graphchi}, Apache Lucene~\cite{bialecki2012apache} and h2 from the DaCapo suite~\cite{blackburn2006dacapo}. We also use microbenchmarks stressing particular aspects of the system to  support our conclusions.

We contribute the following new experimental insights and engineering artifacts.
 First, we evaluate GC activities in our target applications to determine which GC phases are the best candidates for offloading to the integrated GPU. Although the impact of GC on application performance has been extensively analyzed in previous studies~\cite{gidra2011assessing,blackburn2006dacapo,gidra2013study,gidra2015numagic}, none of them have been done in the context of our research question on our integrated system discussed in Section~\ref{subsec:setup}. 
 
 Second, we create a framework for offloading GC tasks on integrated CPU-GPU systems from within the JVM. Our framework is based on the Heterogeneous System Architecture (HSA~\cite{hsa-foundation}) and we describe it in Section~\ref{sec:HSAgarbage}. 
 
 Third, we use our framework to offload the major identified phases of GC to the GPU, analyze the performance (Sections~\ref{sec:copying} and~\ref{sec:marking}), and present our findings, showing that there is generally no observable benefits for accelerating GC on current HSA-capable integrated CPU-GPU systems. We conclude with the discussion of hardware and software limitations that led to our observations.

\section{\textbf{Background}}

\subsection{\textbf{GPUs and Integrated Systems}}
Integrated CPU-GPU systems provide a unified memory view from both the CPU and the  GPU. This unification, at least in theory, allows for seamless processing of complex pointer-based data structures from both processors. In systems where the CPU and GPU do not share the virtual memory address space, pointers to memory allocated by the CPU cannot be used on the GPU. For example, data structures containing pointers to objects cannot be simply referenced on the GPU without deep-copying the objects from the CPU to the GPU and back to the CPU when needed. With shared virtual memory, the CPU can simply pass the same pointer to the data structure and the GPU will be able to seamlessly traverse the structure and its contained pointers and objects.

Memory unification between the CPU and GPU, however, has software and hardware limitations, and hence adapting applications to run on such integrated systems can be challenging. For example, it is still impractical to implement a full cache coherency scheme between the CPU and GPU since this can be very costly and would require complex architectural design choices. Furthermore, the existing heterogeneous programming frameworks such as CUDA and OpenCL provide a variety of memory management methods that interact with the driver and hardware in poorly understood ways, adding programming complexity and limiting performance~\cite{dashti2017analyzing}.

\subsection{\textbf{Garbage Collection}}
A garbage collector is the component in managed runtimes that is responsible for memory allocation and deallocation. The collector tracks down unused objects and deletes them without programmers' intervention. In contrast, in non-managed languages such as C/C++, programmers must explicitly delete objects when they decide that the objects are no longer needed. Managed runtime environments are becoming increasingly popular as they eliminate common memory bugs. However, GC may have exceedingly higher overheads than explicit memory management~\cite{blackburn2004myths,hertz2005quantifying,hertz2005garbage}. Many of the current GC implementations  use generational or regional schemes where they divide the heap into multiple areas to reduce the overhead of collection. One of the major costs of these \emph{tracing} collectors is having to traverse all live objects and their descendants during a collection. Furthermore, many ``big data'' applications retain lots of live data for long periods of time, contradicting the  hypothesis that most objects die young~\cite{lieberman1983real}. With many live objects, the young generation  will quickly fill up and cause frequent collection cycles, which involve heap traversal and object copying and can be costly  compared to explicit memory management. The activities of GC can account for a significant portion of application's execution time, specially for big data systems~\cite{yuan2016spark}.

Different garbage collectors apply different techniques to have as little effect as possible on applications. The main aim of all techniques is to maintain application throughput and reduce \emph{pause times}, the periods of times in which application threads are interrupted. Some collectors use the \emph{stop-the-world} strategy, where they let an application run uninterrupted outside of collection cycles, but during the collection cycle all application threads are halted. A garbage collector may use multiple cooperating threads to perform its work during the collection cycle, which gives it the property of being a \textit{parallel} collector as opposed to being a \textit{serial} one. Using multiple threads aims at reducing the pause times during collections. However, this involves trade-offs as the GC threads need to coordinate their work.

Other techniques for reducing GC overhead include dividing the heap into multiple regions and keeping statistics about each, so that garbage can be quickly collected from regions with the most garbage; the \textit{Garbage First} or G1GC collector~\cite{detlefs2004garbage} uses such a technique. The downside of this method is that it requires a large memory footprint for maintaining metadata about regions and must track cross-region memory references.

Some collectors, including G1GC, also employ \textit{concurrent} phases for some of their activities.  Concurrent phases allow the collector to perform its work while the application threads are executing. This adds complexity since both the GC threads and application threads may potentially modify the same objects concurrently.  Therefore, concurrent collectors must use techniques such as \textit{write barriers}, or \textit{read/load barriers} to ensure correctness. These barriers are snippets of code that are added to the application code by the JIT (Just In Time) compiler whenever a read or write reference is encountered. They ensure atomicity of memory accesses and perform bookkeeping.

Regardless of how a garbage collector is designed, the main activities in any collector are essentially graph traversals (with possibly concurrent node modifications) and bandwidth-constrained memory operations for copying, moving, and/or compacting memory objects on the heap. GPUs have relatively large memory bandwidths and their architectures are designed to hide memory latency with extreme parallelism. Such observations arguably make GPUs a potential target for running GC. Furthermore, hardware systems are increasingly moving toward integrating GPUs on chip along with a multicore CPU to eliminate the high costs of copying data from/to CPU/GPU. Although these integrated GPUs have less memory bandwidth and are less computationally capable than discrete GPUs, not needing to copy data can offer both performance and programmability advantages. 
\section{\textbf{Analyzing the Overhead of Garbage Collectors}}
\label{sec:analysisGC}

In this section, we measure the effects of GC on our target applications running only on the CPU. We describe our experimental setup in section~\ref{subsec:setup}, and analyze the overhead of GC in section~\ref{subsec:motivation}. We also break down the major phases of GC to show that object copying and marking are dominant.

\subsection{\textbf{Experimental Setup}}
\label{subsec:setup}
In all of the experiments in this paper we use an AMD Carrizo A10-9600P APU running Ubuntu Linux 18.04. It is a 7th generation Bristol Ridge series APU with 4 CPU cores (two Excavator modules) clocked at 2.4 - 3.3 GHz. The L2 cache is 2 MB. The integrated GPU is a Radeon R5 composed of 6 Compute Units (CU) and clocked at 720 MHz. The total memory capacity of the system is 12GB.

For our case study we use Apache Spark (spark-sql-perf), Apache Lucene (text search using a Wikipedia input file), GraphChi (Pagerank using Twitter input graph), and h2 from Dacapo. We analyzed many more applications and benchmarks, but identified these four as the most affected by GC and showing a variety of nuance in how GC can affect performance. We use the following standard garbage collectors that are available on modern OpenJDK versions: 
\begin{itemize}
    \item SerialGC: A generational stop-the-world throughput-oriented collector. This is an old GC, but it serves a useful reference point. 
    \item ParallelGC: A generational stop-the-world throughput-oriented collector that uses multiple threads during some collection phases. This is the default collector in Java 8~\footnote{Java 8 is still widely used and is maintained until the end of 2030. Furthermore, in this work we evaluate GPU offloading of copying and marking which are major phases in almost all modern garbage collectors, so we believe our results are applicable to different GCs in any Java version.}.
    \item G1GC: A generational, parallel, regional, and mostly-concurrent garbage collector. It is mostly-concurrent since it only employs concurrency with application threads during marking. Other phases in this collector are stop-the-world. This collector has become the default garbage collector since Java 9.
\end{itemize}

We use a number of configurations for each collector depending on the experiment, as specified in the following sections.

\subsection{\textbf{Running GC on the CPU}}
\label{subsec:motivation}

\begin{figure*}[ht]
    \centering
    \begin{subfigure}[b]{0.48\textwidth}
        \includegraphics[width=\textwidth]{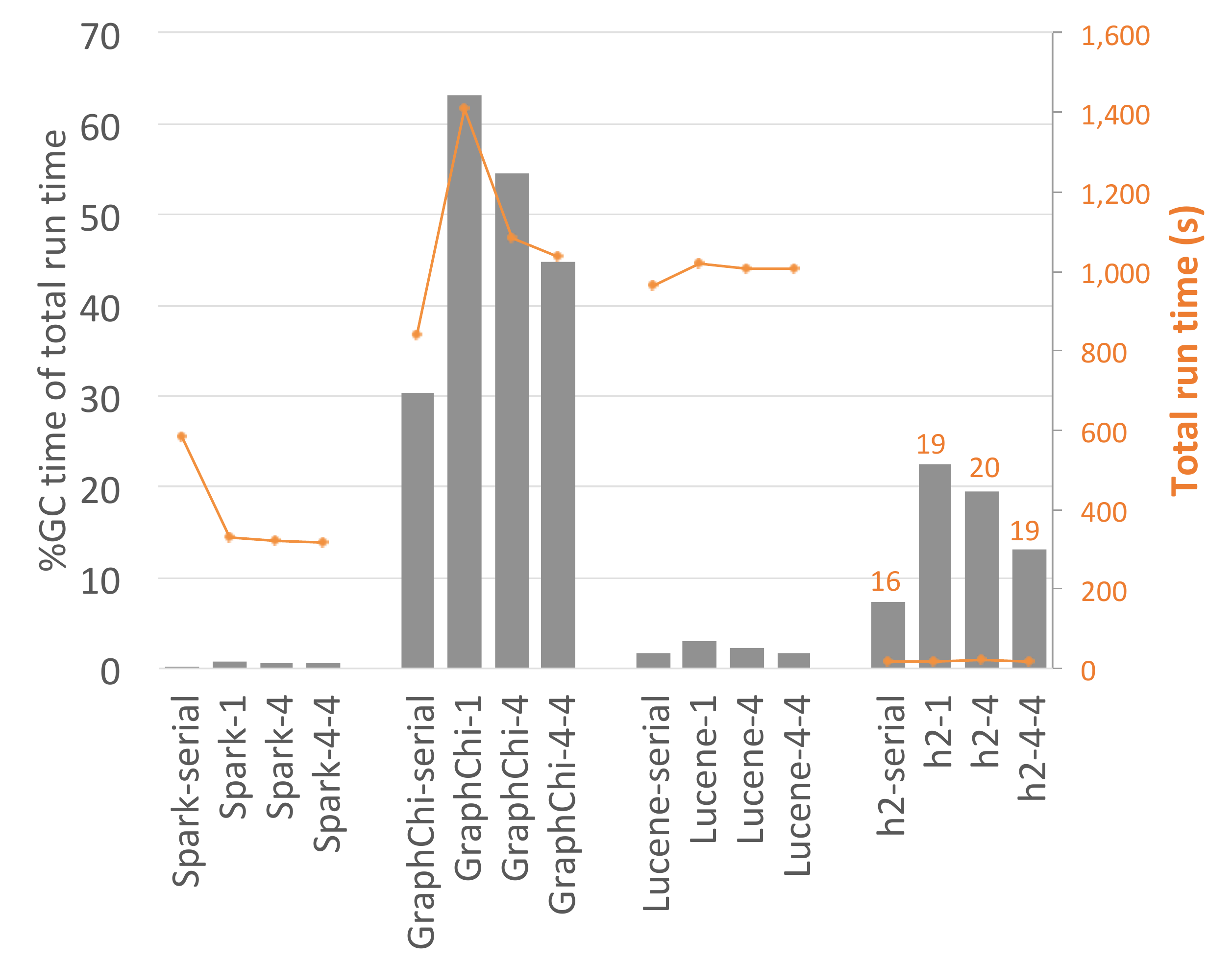}
        \caption{Total run time (orange) and \%GC time (gray).}
        \label{fig:summary-chart1}
    \end{subfigure}
    ~
    \begin{subfigure}[b]{0.48\textwidth}
        \includegraphics[width=\textwidth]{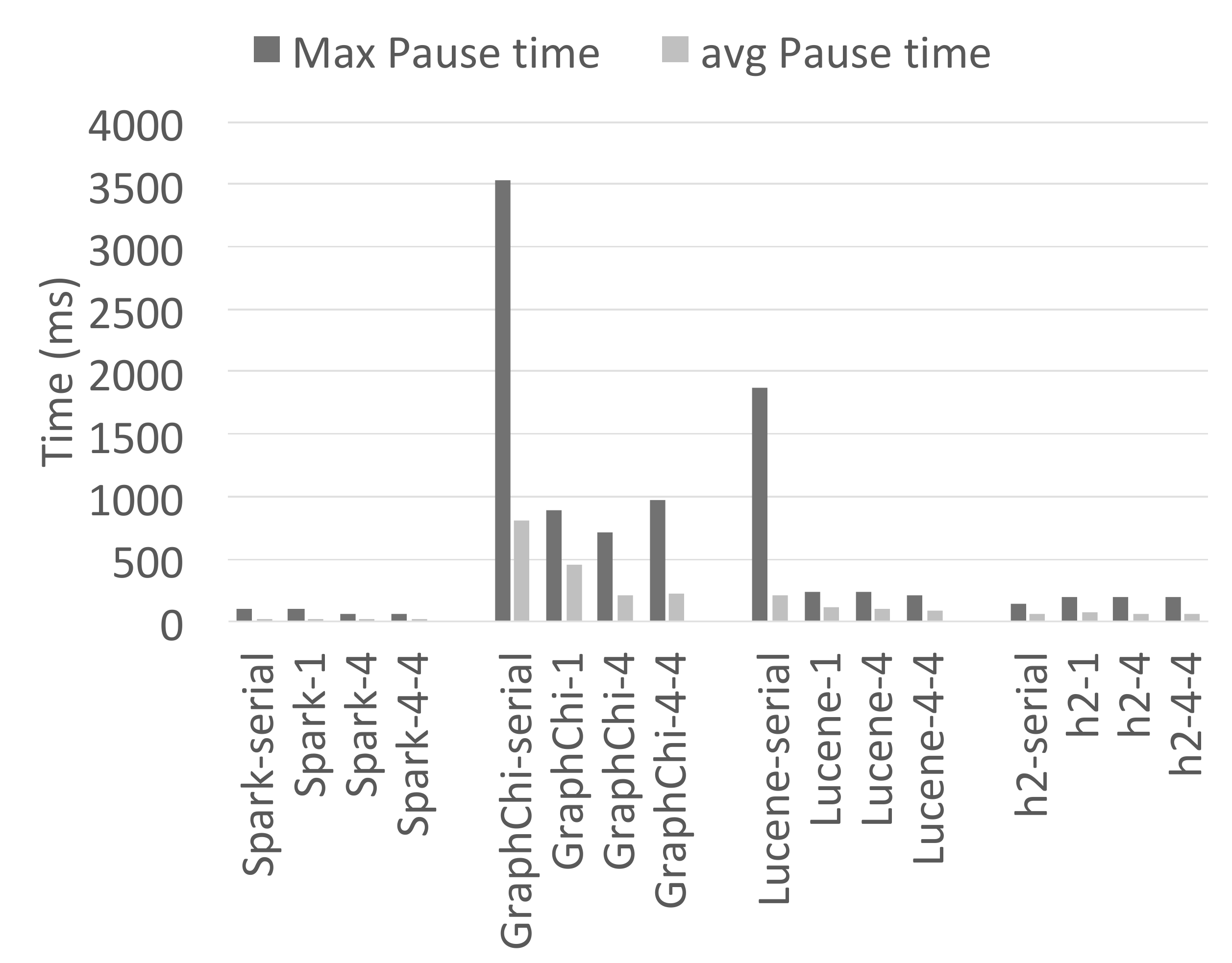}
        \caption{Max. and Avg. pause times.}
        \label{fig:summary-chart2}
    \end{subfigure}
    \caption{(APP)-serial is the app running with SerialGC .(APP)-1 is the app running with G1GC using 1 GC thread. (APP)-4 is the app running with G1GC using 4 GC threads. (APP)-4-4 is the app running with G1GC using 4 GC threads and 4 Concurrent Marking threads.}
    \label{fig:summary-charts}
\end{figure*}

\subsubsection{\textbf{First Experiment: Effects of GC on applications}}
How much GC affects applications depends on many factors, including the behavior of the application itself. If it rarely allocates any objects in the heap, there will hardly be any GC cycles that might delay its execution. The impact also varies depending on how long the application retains allocated objects. Furthermore, the time spent on GC is not always a direct predictor of GC-related performance degradation, as we will see below.

Figure~\ref{fig:summary-chart1} shows the percentage of the application time spent on GC. Each application (APP) is run under four GC configurations: (APP)-serial uses SerialGC .(APP)-1 uses G1GC using one GC thread. (APP)-4 uses G1GC with four GC threads and one marking thread. (APP)-4-4 uses G1GC using four GC threads and four concurrent marking threads. All  configurations use four application threads to perform the actual work.

We see in figure~\ref{fig:summary-chart1} that GC affects applications differently, which has already been demonstrated by many previous works such as~\cite{detlefs2004garbage,blackburn2004myths,hertz2005quantifying,marlow2008parallel}. For some applications (i.e. Spark and Lucene), GC does not constitute a significant portion of the execution time. This, however, doesn't mean that these applications are not affected by the GC implementation since there are many other factors such as barriers and lock contention. For example, we found out that when an application runs with multiple mutator threads, it is possible that these threads compete for lock-based new memory allocations on some shared regions. This is true for Spark with Parallel and Serial GC. We also noticed that when the \emph{thread-local allocation buffers} (TLABs) become full and new ones need to be acquired from the JVM, the contention for acquiring new buffers can negatively affect performance. These effects are sometimes not counted as GC overhead, since they may be part of the JIT-generated code. 

Furthermore, although SerialGC might appear to provide a low-overhead and high-throughput alternative to concurrent or parallel collectors, it causes massive pause times. Since the serial collector must stop all application threads to clear the garbage and only utilizes a single CPU core to do the collection, applications will become unresponsive during this period. This may be unacceptable to interactive applications or those with soft-real time requirements. Figure~\ref{fig:summary-chart2} demonstrates these effects by showing the maximum and average pause times for the applications.

\begin{figure}
\centering
\begin{tabular}{cc}
  \includegraphics[scale=0.15]{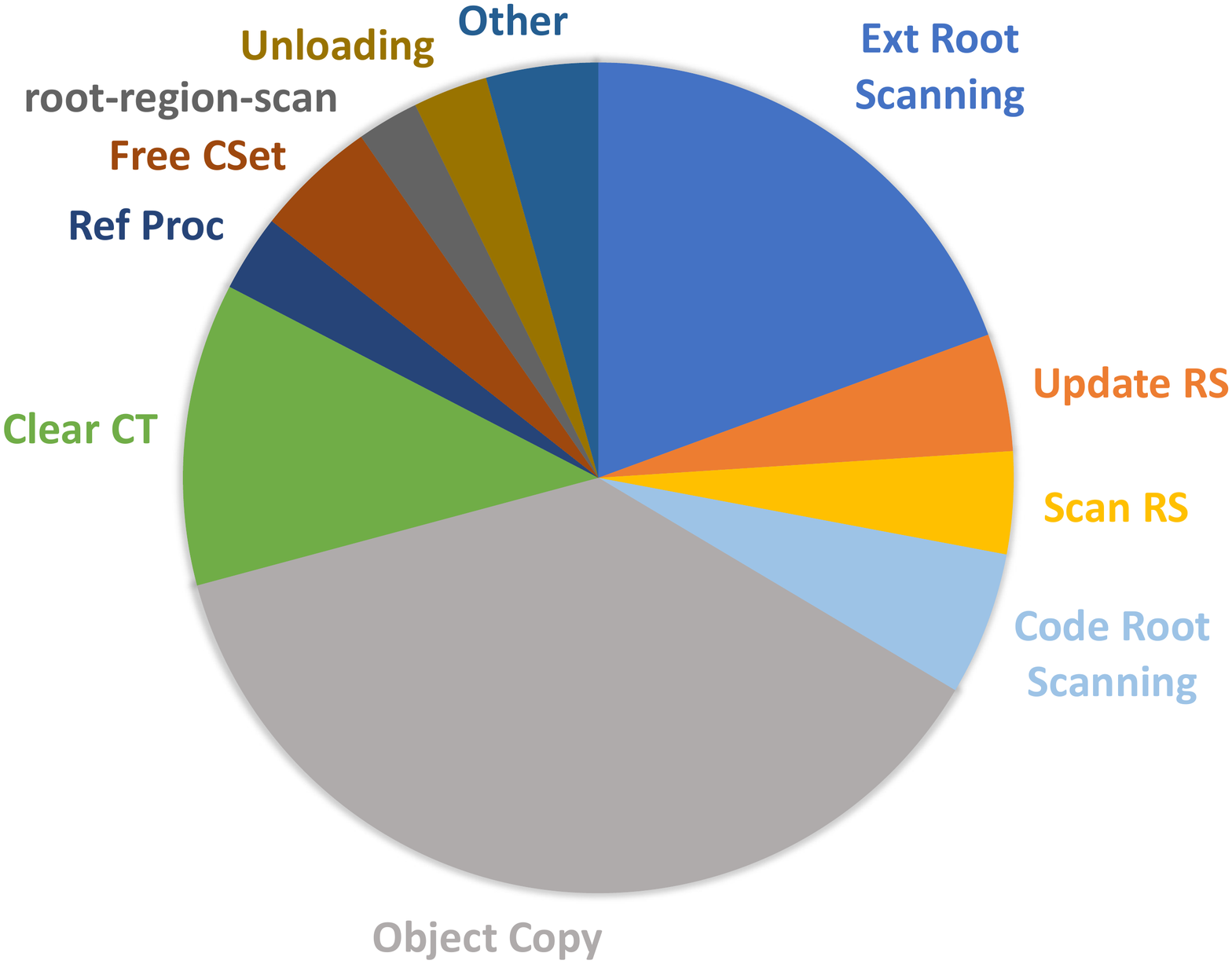} &   \includegraphics[scale=0.15]{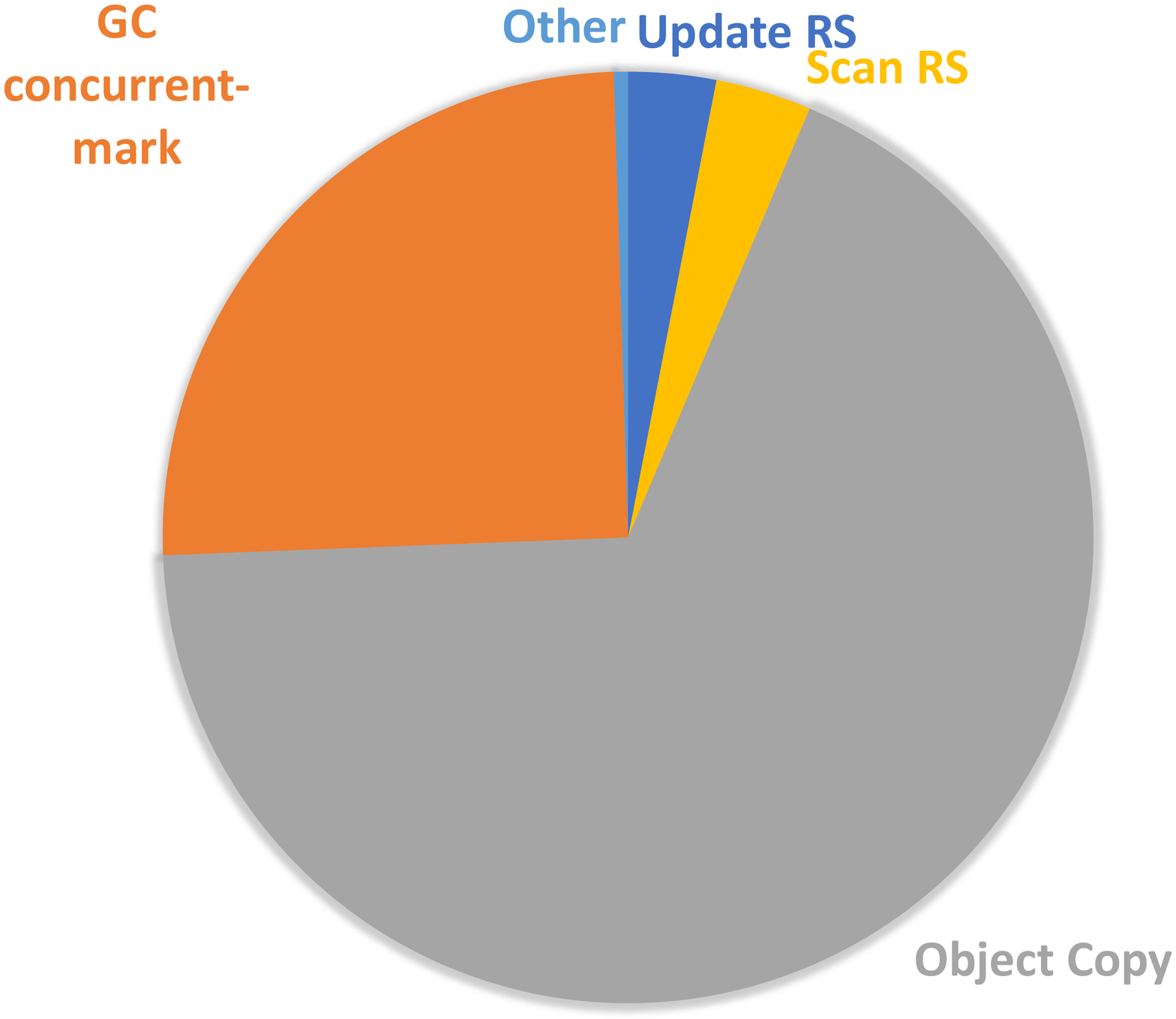} \\
(a) Spark & (b) GraphChi \\
 \includegraphics[scale=0.15]{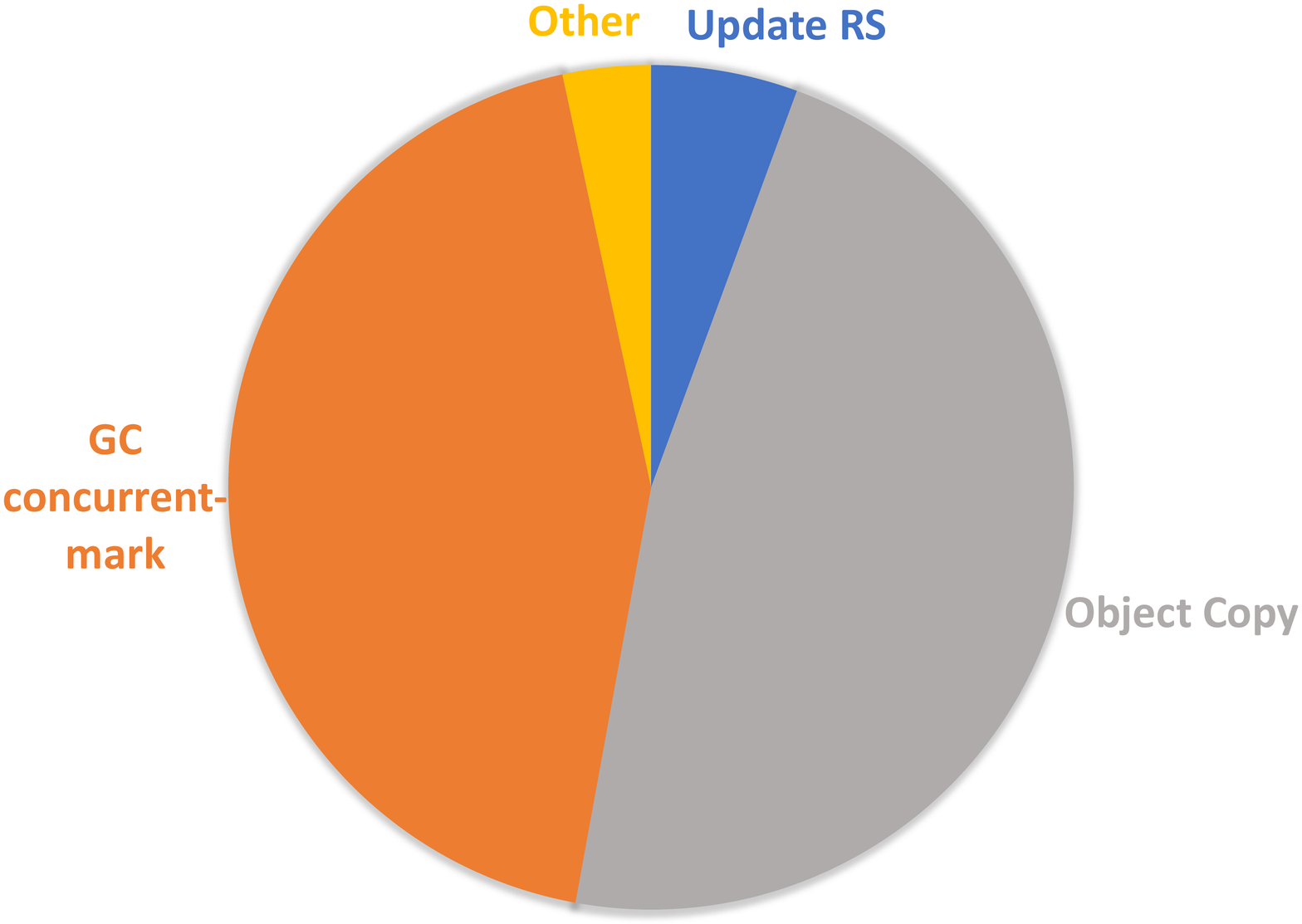} &   \includegraphics[scale=0.15]{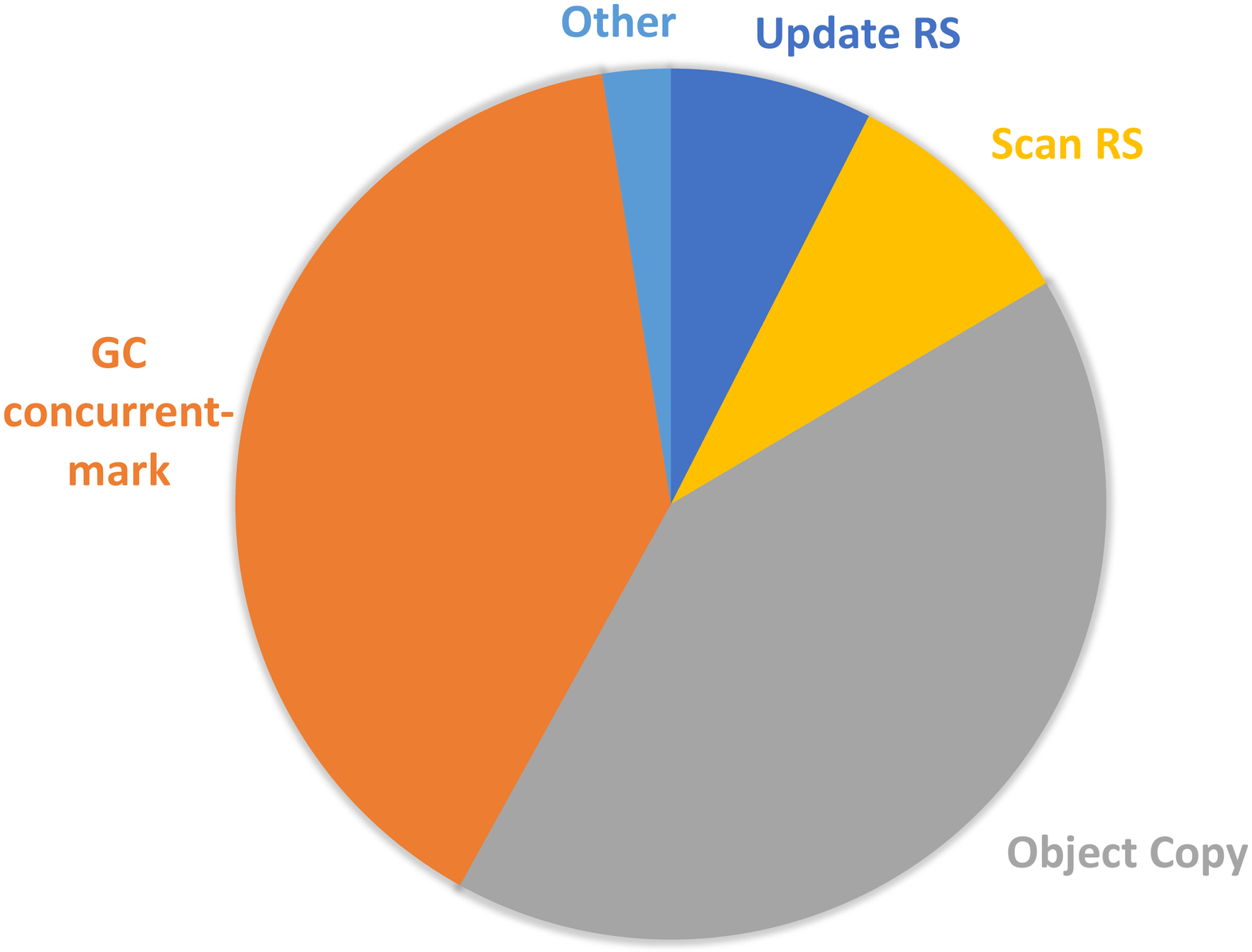} \\
(c) Lucene & (d) h2 \\
\end{tabular}
\caption{Breakdown of GC phases. RS: Remembered Set. CT: Card Table. Ref Proc: Reference Processing. CSet: Collection Set. Ext: External}
\label{fig:GCbreakup}
\end{figure}

\subsubsection{\textbf{Second Experiment: Breakdown of GC phases}}
Garbage collection consists of several phases and the impact of each phase differs depending on the behaviour of a given application. In this section we break down the overhead of these  phases. We  focus only on G1GC here, since major phases are similar in other garbage collectors, and for brevity we show the break-down of APP-1 experiments. Running multiple GC and marking threads impacts the total time spent in the phases, but not their relative contribution to the total overhead.

Figure~\ref{fig:GCbreakup} shows the breakdown of the GC overhead. The main observation from these charts is that object copying and concurrent marking are the  dominant phases for all GC-impacted applications. The only exception is Spark. The behavior of Spark is that it allocates huge amounts of data and has a very high allocation rate, yet it does not retain most of this data so almost all the allocated objects become garbage instantaneously in the young generation. Since generational collectors will allocate objects in the young generation first, and many of these objects will quickly be dead, the phases of garbage collection will quickly terminate without much work (i.e. there is not much traversing of the live object graph).

Also, we showed in the previous experiment that GraphChi, which retains a large portion of its allocated objects, is heavily impacted by GC, taking up to 63\% of its runtime (Figure~\ref{fig:summary-chart1}). As Figure~\ref{fig:GCbreakup}b shows, more than 70\% of total GC overhead in GraphChi is due to object copying, which is done in a stop-the-world fashion. Live objects in G1GC are compacted by copying them into new regions. The more live objects there are, the more copying will take place. 

Concurrent marking is where GC marking threads will compete with application (mutator) threads as they mark a (possibly changing) live object graph. Both the GC threads and application threads might simultaneously modify object references on the heap; GC threads are traversing and marking objects as "live" whereas application threads might be creating or deleting references from the same objects. This might introduce overhead as the correctness of the object graph has to be maintained. Although, G1GC uses concurrent marking, the marking phase can account for a significant portion of the total GC time. Different applications vary in how they are affected by GC, but overall, copying and marking phases dominate.


\section{\textbf{Using HSA for Garbage Collection Acceleration}}
\label{sec:HSAgarbage}
To be able to modularize and easily port GC tasks to run on the GPU, we created an HSA~\cite{hsa-foundation} module that is integrated within the Hotspot OpenJDK code. The module connects the JVM with the HSA framework and provides a simple API to the JVM to launch GPU kernels. Using this module, programmers write GPU kernels for any task they wish to offload to the GPU. The GPU kernels are written in OpenCL (HSA uses OpenCL as the language for writing GPU tasks) and they can be launched from anywhere in the JVM code using an API call.

In the following sections, we use this HSA module to analyze the object copying and marking phases (the two dominant phases as discussed in the previous section) and study the potential of accelerating them on the integrated GPU.

\subsection{\textbf{Investigating the Offloading Potential of Object Copying}}
\label{sec:copying}


\subsubsection{\textbf{CPU vs. GPU Object Copying}}

\begin{figure}[t]
\centering
\includegraphics[width=\linewidth]{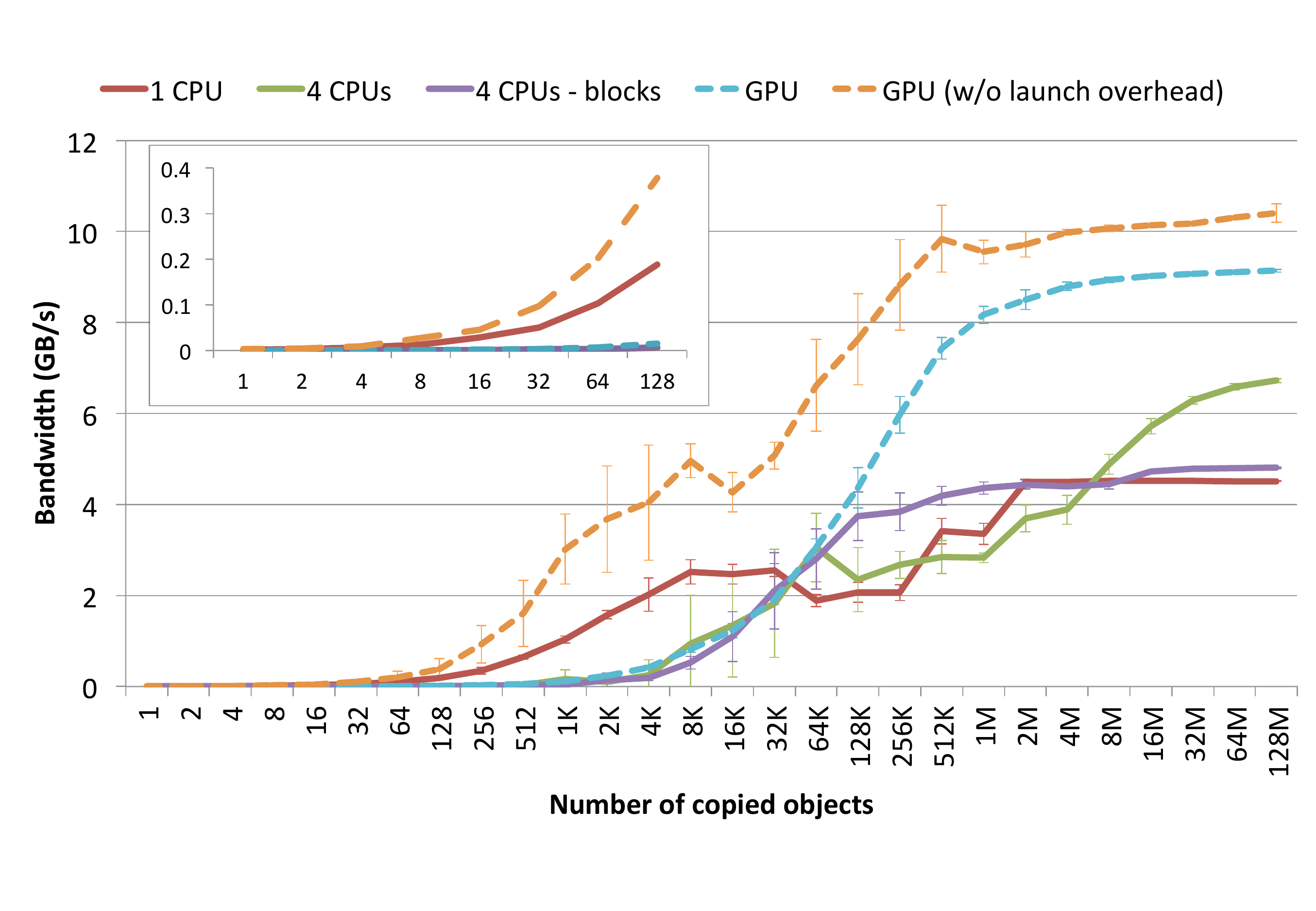}
\caption{CPU vs. GPU object copy bandwidth}
\label{fig:bandwidth}
\end{figure}

We first look at the cost of copying objects using the CPU and the GPU. In this experiment (figure~\ref{fig:bandwidth}), we compare the bandwidth that can be achieved when we copy an increasing number of objects. In this experiment, all data exist in main memory, so there are no disk accesses or page faults that affect the results. Each single object has a size of 8 bytes; when we copy 128K objects, for example, we are copying 1MB of total data. We compare five cases: 1) A single CPU thread copies the objects using \texttt{memcpy}. It copies all of the objects at once (a single \texttt{memcpy} call is issued to copy any number of objects).  2) Four CPU threads split the total number of objects among them and perform the copying with \texttt{memcpy}. 3) Four CPU threads split the total number of objects among them but further divide their portion into multiple smaller blocks that each thread will copy in a for loop (this is 4 CPUs-blocks in the figure). This way each \texttt{memcpy} call will copy a smaller buffer in each loop iteration instead of one big portion which can sometimes be more efficient. 4) The GPU performs the copying using direct assignment (\texttt{to\_buffer[i] = from\_buffer[i]}) where each entry in the array is of size equal to 8 bytes (\texttt{uint64t\_t}). 5) The GPU performs the copying but we exclude the kernel launch overhead. This case could represent future systems where the cost of launching kernels is negligible. 

For GPU copying, the GPU kernel is launched with as many threads as there are objects to copy; hence, the launched kernel will utilize up to all of the 6 Compute Units (CU) available. There are 384 SIMD (Single Instruction Multiple Data) lanes within these CUs, and these lanes can be used simultaneously. Each GPU thread in a kernel copies 8 Bytes of data. So, if we launched, for example, 1024 GPU threads, we will copy 8KB of data simultaneously (of course, we are still limited by the physical SIMD lanes available, but all of the threads will be ready to run hiding the memory latency of accesses~\cite{bauer2014singe}).



There are two main insights which can be drawn from the results in figure~\ref{fig:bandwidth}. First, if we do not consider the cost of launching GPU kernels, using the GPU to perform object copying provides the best bandwidth utilization. Object copying consists of simple operations for reading from and writing to memory addresses. GPUs excel at hiding the latency of memory accesses and can launch thousands of threads to perform the copying. CPUs, however, have to rely on limited-width SIMD instructions (when using \texttt{memcpy}) which cannot seem to fully utilize the available bandwidth when copying data.

The second insight is that the GPU (including the launch overhead) only becomes better than the CPU when copying many objects (and hence large data sizes). The GPU exceeds single CPU thread copying at 64K objects (512KB), and it only becomes better than four CPU threads at 128K (1 MB). Although the GPU might not offer a substantial bandwidth advantage except for relatively larger data copies, it could still be used as an alternative processor to perform the copying. Therefore, having a GPU on board could still be favourable as the CPU becomes available for other more complex computations.


\subsubsection{\textbf{Offloading Promotional Object Copying to the GPU}}

To further analyze offloading object copying, we targeted heap object copying during the promotion phase of a parallel generational garbage collector. The promotion phase copies all live objects from the young generation to the old generation. Object copying from young to old generation occurs when the young generation becomes full of live objects that cannot be collected, so they need to be copied to the old generation. We modify OpenJDK's Java 8 Hotspot parallel garbage collector to use our HSA framework for launching GPU kernels that perform object copying. We call this scheme \textit{ParallelGPUCopy}.

Our HSA framework allows us to target any garbage collector, but for our tests in this section we augmented ParallelGC with HSA for three main reasons. First, this collector is the default collector in Java 8, so we can compare our results when offloading to the GPU to the default collector performance. Second, ParallelGC is a stop-the-world collector, which makes the implementation much easier. We will discuss later in section~\ref{sec:marking} some of the issues with concurrent GCs. Finally, we targeted ParallelGC since it allows using multiple GC threads, so we can verify that our HSA module enables concurrent kernel launches from multiple GC threads. This verification ensures that the implementation is thread-safe and guarantees the ability to launch GPU kernels concurrently from any JVM thread.

We wrote a Java micro-benchmark and run it using our modified JVM (ParallelGPUCopy scheme) to test the performance impact of offloading the object copying tasks. This is similar to the experiment in figure~\ref{fig:bandwidth} but applied to a real scenario integrating the GPU copying functionality into the JVM using our HSA module. 

In order to create a scenario to test promotional object copying, we need to make sure that the benchmark keeps on allocating objects that remain to stay alive. A simple way to do this is by adding objects into a linked list. The micro-benchmark does this by specifying the total number of objects to allocate and the size of objects. In the tests, a total amount of data equal to 4GB is allocated. So, the linked list will contain 512K objects of size 8KB each, 256K objects of size 16K, 128K objects of size 64K, and so on. The heap object graph in this application is not very parallelizable: the graph looks like a linked list which has minimal parallelization advantage; However the objects will remain alive as they won't be quickly orphaned and hence would continuously be promoted to the old generation. The microbenchmark is intentionally designed to stress such behaviour so that the potential of using the GPU to do the object copying during promotion is analyzed.

\begin{figure}[t]
\centering
\includegraphics[width=\linewidth]{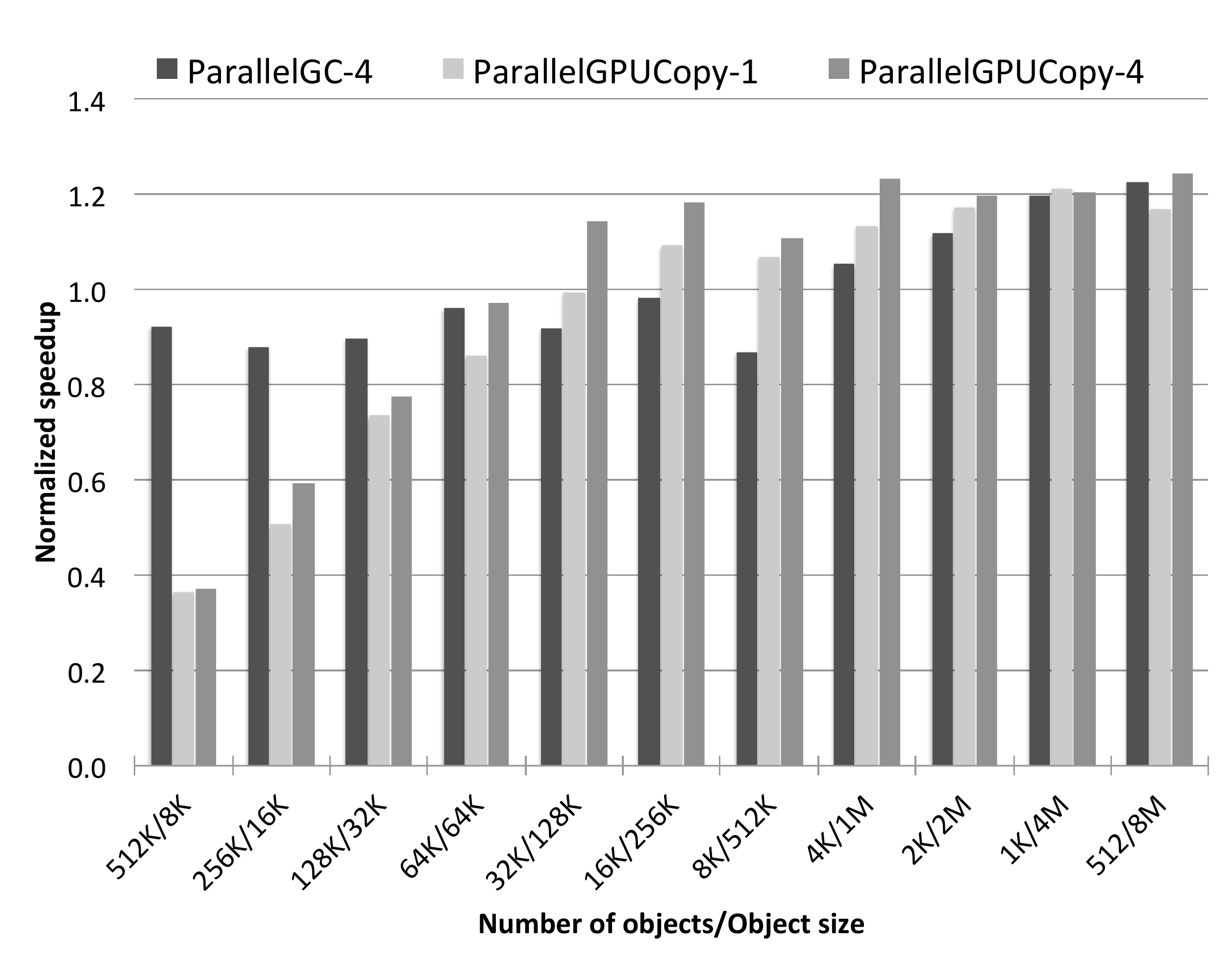}
\caption{Normalized speedup comparing ParallelGC and ParallelGPUCopy for promotional object copying. Each collector is configured to use either 1 CPU thread, or 4 CPU threads. In the case of multiple threads, GPU kernels are launched concurrently from threads.}
\label{fig:GC2}
\end{figure}

We compare four garbage collectors: 1) the baseline is the default single threaded ParallelGC, 2) a 4-thread ParallelGC which only uses the CPU cores, 3) ParallelGPUCopy with a single thread performing copy operations on the GPU, and 4) a 4-thread version of ParallelGPUCopy where multiple threads launch GPU kernels concurrently. Figure~\ref{fig:GC2} shows the speedup of the collectors over the baseline. The x-axis shows both the total number of object and the size of each object. The total number of objects is 4GB divided by the size of the object (we allocate a total of 4GB in each case). For smaller object sizes, the GPU is inferior as was demonstrated in the earlier experiments. As the number of objects to be copied exceeds 32K (128KB), we start to gain performance improvements compared to the baseline single CPU copying.

An important difference between the setup of the experiment here and the previous experiment (figure~\ref{fig:bandwidth}) is that the HSA module is used in isolation outside the JVM in the previous experiment; therefore, there are no other JVM activities which might indirectly affect object copying. Here (figure~\ref{fig:GC2}), the HSA module is integrated within the JVM, and the object copying kernel is called from within the JVM by the GC threads that are performing the promotions of live objects.

Although the trends in both experiments are the same, we notice that there is a slightly lower advantage to GPU copying when applied inside the JVM. The reason for this is subtle and has to do with synchronization. The JVM launches many threads to handle various activities. One of the core threads is the thread responsible for orchestrating a so-called "safe point". A safe point is required, for example, when all application threads need to come to a stop to perform a stop-the-world phase in any garbage collector (this not specific to ParallelGC and is also applicable to G1GC during the young evacuation pause where live objects get copied to other regions). After the GC threads finish their tasks, they have to communicate with the VM thread to indicate that they are done. This synchronization may take an unpredictable amount of time. So even though the GC threads using the GPU for copying may be done sooner than their CPU counterparts, they still have to wait for the VM "safe point" enforcer thread to allow them to proceed, and so the overall advantage of faster copying is dampened.


Although object copying is a major performance bottleneck during GC cycles for ``big data'' applications, as shown in prior work~\cite{bruno2017ng2c} and reconfirmed here, applications need to generate lots of live and relatively large data objects to reap any benefits of GPU copying. One application in our study, GraphChi, possesses this property and in the following section we examine it further.


\begin{figure}[t]
\centering
\includegraphics[width=\linewidth]{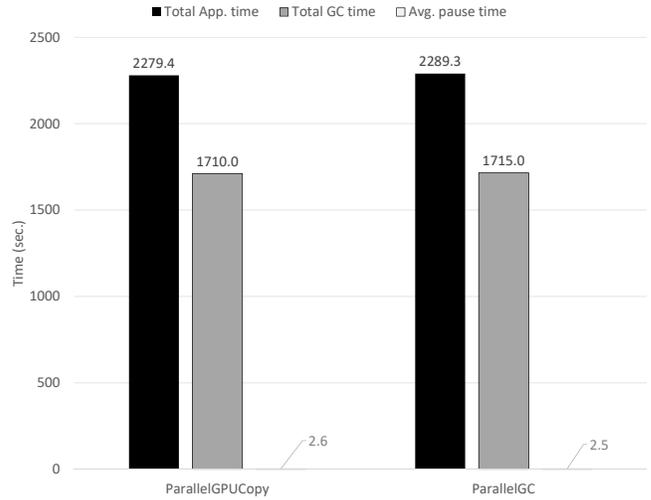}
\caption{Performance of GraphChi with object copying offloaded to the GPU compared to default ParallelGC.}
\label{fig:gpugcgraphchi}
\end{figure}

\subsubsection{\textbf{Further Investigation of GraphChi}}
We further tested GraphChi under ParallelGPUCopy to see if there are any benefits to offloading object copying. GraphChi uses a lot of big objects which makes it a good candidate application to test the offloading of copying to the GPU as opposed to the DaCapo applications. In the promotional object copy code in the JVM, we only use the GPU for copying objects with sizes greater than or equal to 32KB~\footnote{We tested several values for this threshold including 64K, 128K, and 1M, and there wasn't a significant difference in performance.}. So, all smaller object sizes will be copied using the CPU.

When we profiled GraphChi, we observed that more than 5K objects with sizes greater than 32KB are promoted during the execution of the application. The average size of these objects is 3.26MB (GraphChi is memory intensive, so we use a heap size of 6GB in this experiment to pressure the JVM to start GC cycles). Therefore, such statistics fit well for having the copying offloaded to the GPU. However, the results shown in figure~\ref{fig:gpugcgraphchi} do no show any substantial benefits for such offloading. GraphChi spends a lot of time in GC, with object copying being the most significant portion of its time. From figure~\ref{fig:bandwidth}, we see that we can copy object sizes of around 3MB at 6GB/s using the GPU compared to 4GB/s using 4 CPUs. So, theoretically (not considering any other JVM side effects), we should expect to speedup the copying of these large objects at a rate of 1.5x. However, this improvement of object copying time does not translate into any significant total run or pause time reduction because of the JVM synchronization we described earlier.

\subsection{\textbf{Investigating the Offloading Potential of the Marking Phase}}
\label{sec:marking}

\begin{figure*}
\centering
\begin{tabular}{cc}
  \includegraphics[scale=0.3]{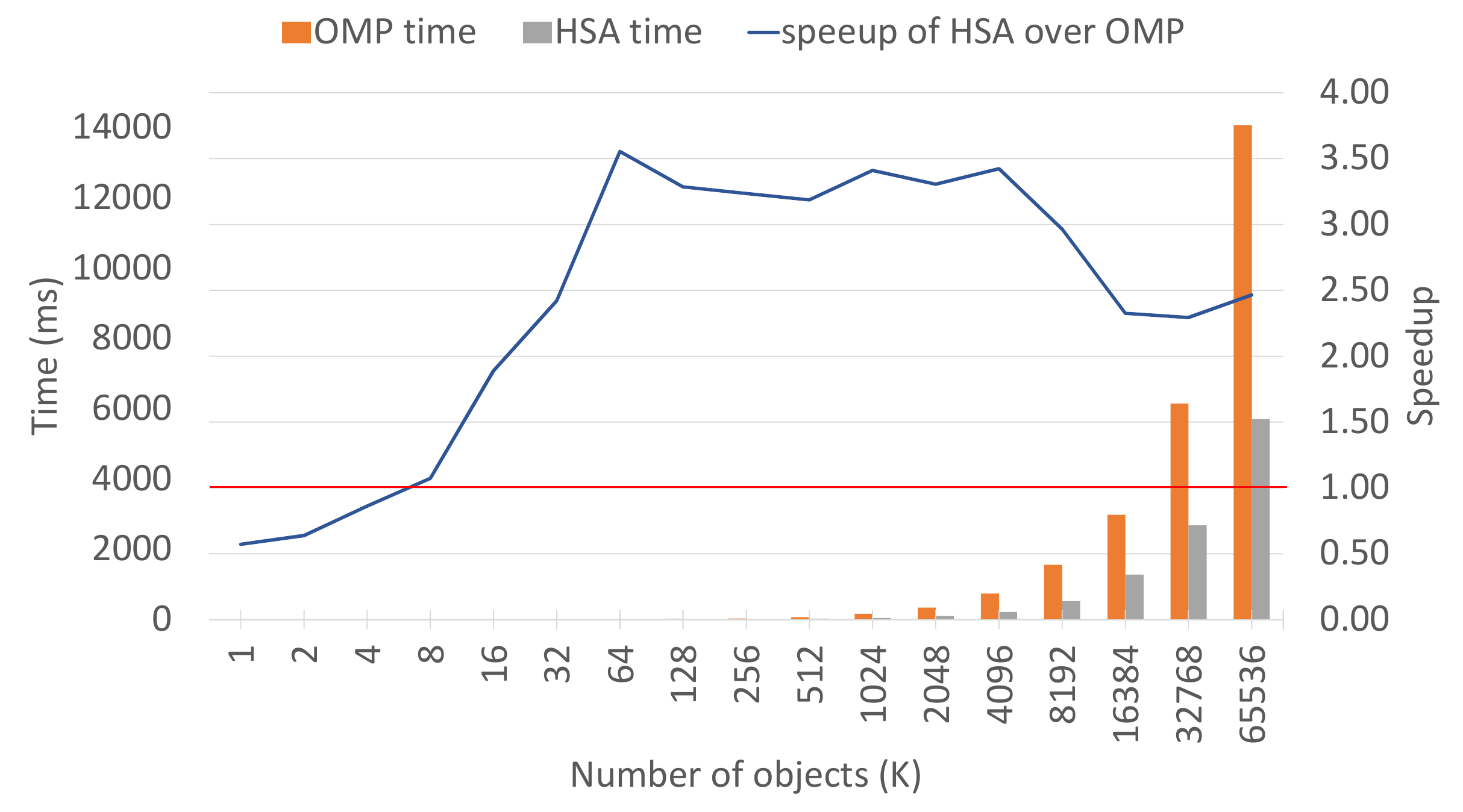} &   \includegraphics[scale=0.3]{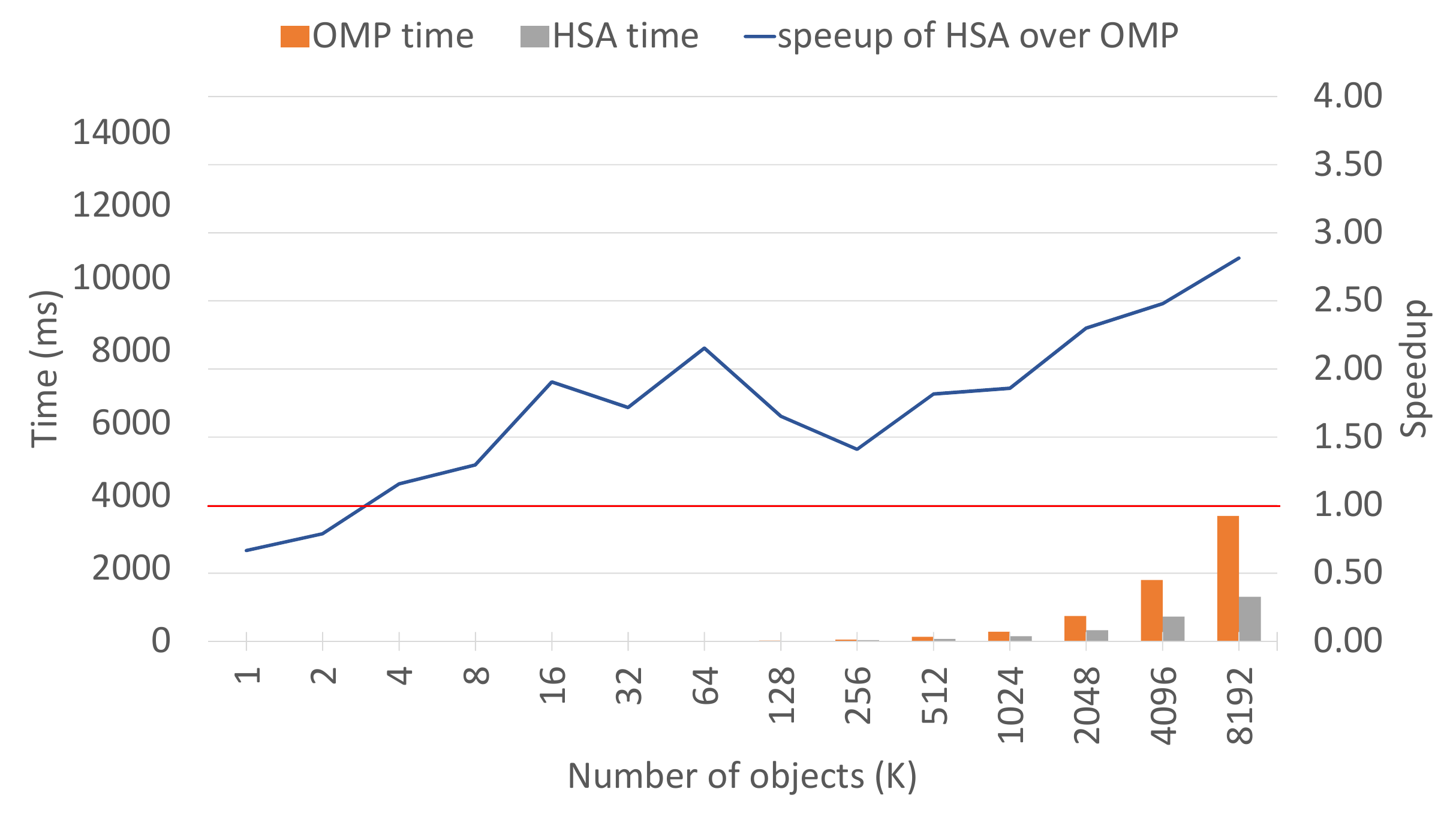} \\
(a) X = 8 & (b) X = 40 \\
 \includegraphics[scale=0.3]{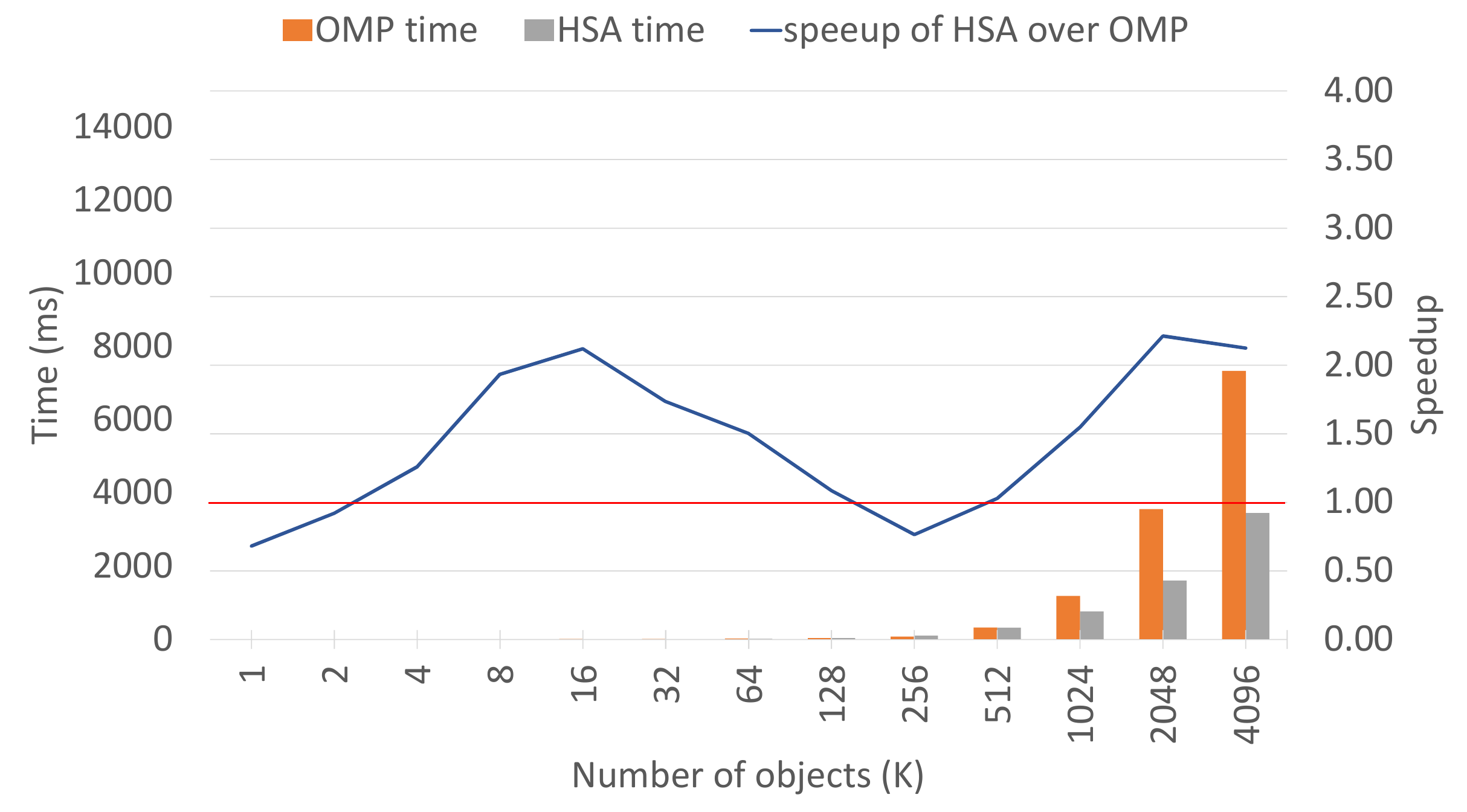} &   \includegraphics[scale=0.3]{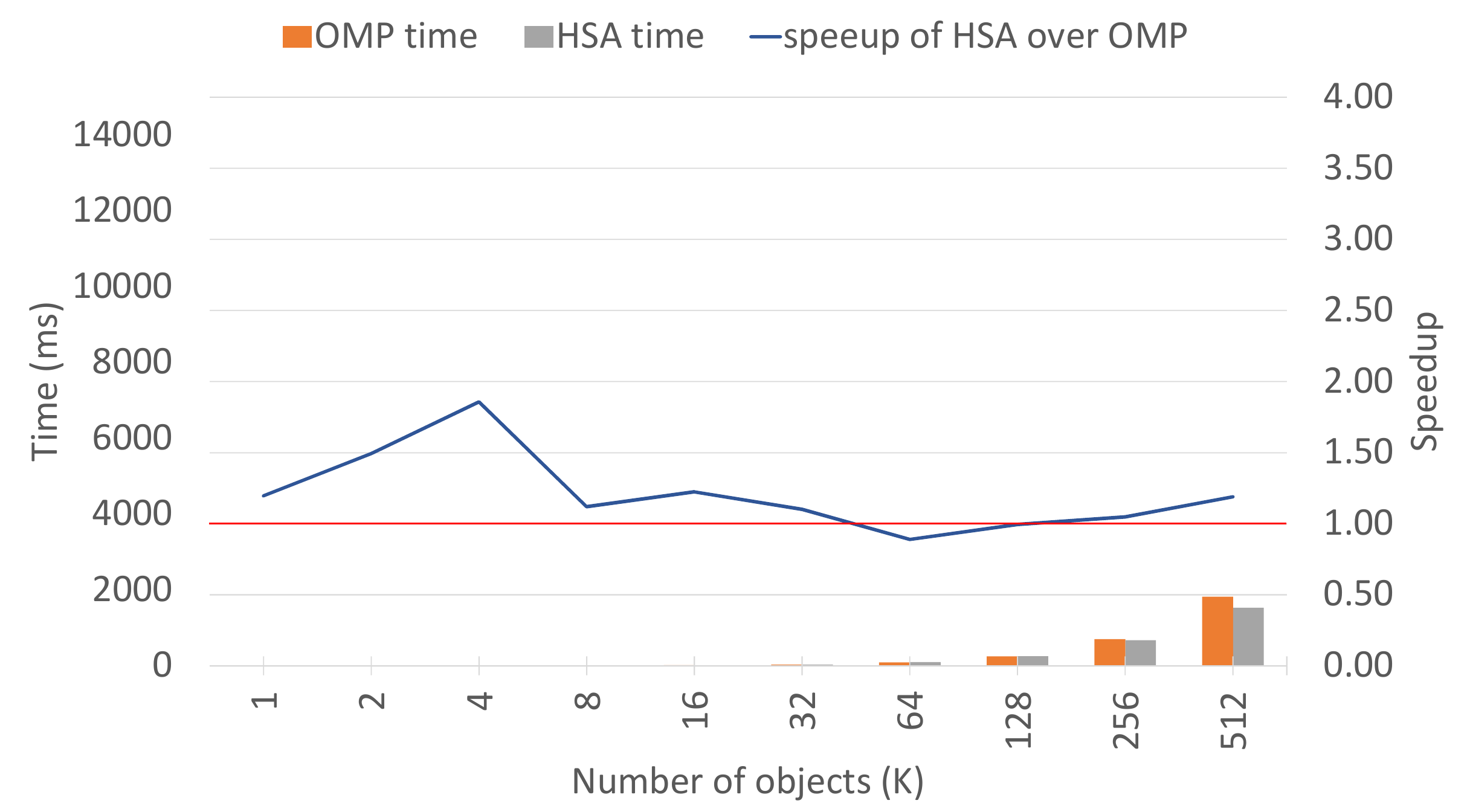} \\
(c) X = 200 & (d) X = 1000 \\
\end{tabular}
\caption{The time to traverse a heap object graph with increasing number of objects. The shape of the graph is randomly constructed with an average number edges = X. Hence, each object will have X links on average to other random objects.}
\label{fig:markingGPU}
\end{figure*}

Marking is the process of traversing the live heap object graph from the roots and setting a bit to identify that an object is "marked". The object graph may keep on evolving as new objects are created, or new links between objects are modified. Marking can either be done in a stop-the-world fashion or concurrently with application threads. Concurrent marking is more complex as application threads may continuously mutate the shape of the object graph by adding, modifying, or removing references to and from objects.

GPUs have been demonstrated to work well with some graph algorithms such as breadth-first search (BFS). Marking is essentially a BFS process on a possibly morphing graph structure. In order to assess the practicality of offloading marking to the GPU using our HSA framework, we used microbenchmarks to model object graphs with varying shapes. We first explore stop-the-world marking and then concurrent marking. 

\subsubsection{\textbf{Stop-The-World Marking}}
Figure~\ref{fig:markingGPU} shows the results from performing the marking phase on four different graph heap shapes represented by a variable 'X' that indicates the average number of edges among nodes (or objects). The shapes range from sparsely distributed objects with few connections (less marking work) to more densely packed connections among objects (more marking work). These tests emulate a stop-the-world marking since there are no mutator threads modifying objects. Hence, the assumption in these tests is that the garbage collector is the sole entity modifying (marking) the graph. The shapes in each example in the figure are static and do not change throughout the execution of the marking phase. However, looking at the different shapes gives an estimate on how good offloading the marking phase is. In these experiments, two variables control the shape of the object graph. The first variable is the total number of objects (nodes) that will be created. The second variable is the connectivity level between these objects, or the average number of links (edges) between objects. The object graph is constructed randomly using these two variables, and the result is a random graph shape which represents a live object graph that is traversed from the roots.

We compare the performance of our HSA framework to an OpenMP version performing the heap traversals on the CPU using all available cores. The results show that depending on the object graph shape, a potential speedup of 3.5x is attainable by offloading the marking phase to the GPU. Of course, as the graph becomes denser and less parallelizable, this potential diminishes as can be seen in figure~\ref{fig:markingGPU}d. Real-world applications produce object graphs of varying dynamic shapes with a changing level of parallelizing potential. So, our results show an upper bound on the maximum attainable speedup on the integrated system used in this paper.

\subsubsection{\textbf{Concurrent Marking}}

\begin{figure*}[!htbp]
\centering
\begin{tabular}{cc}
  \includegraphics[scale=0.3]{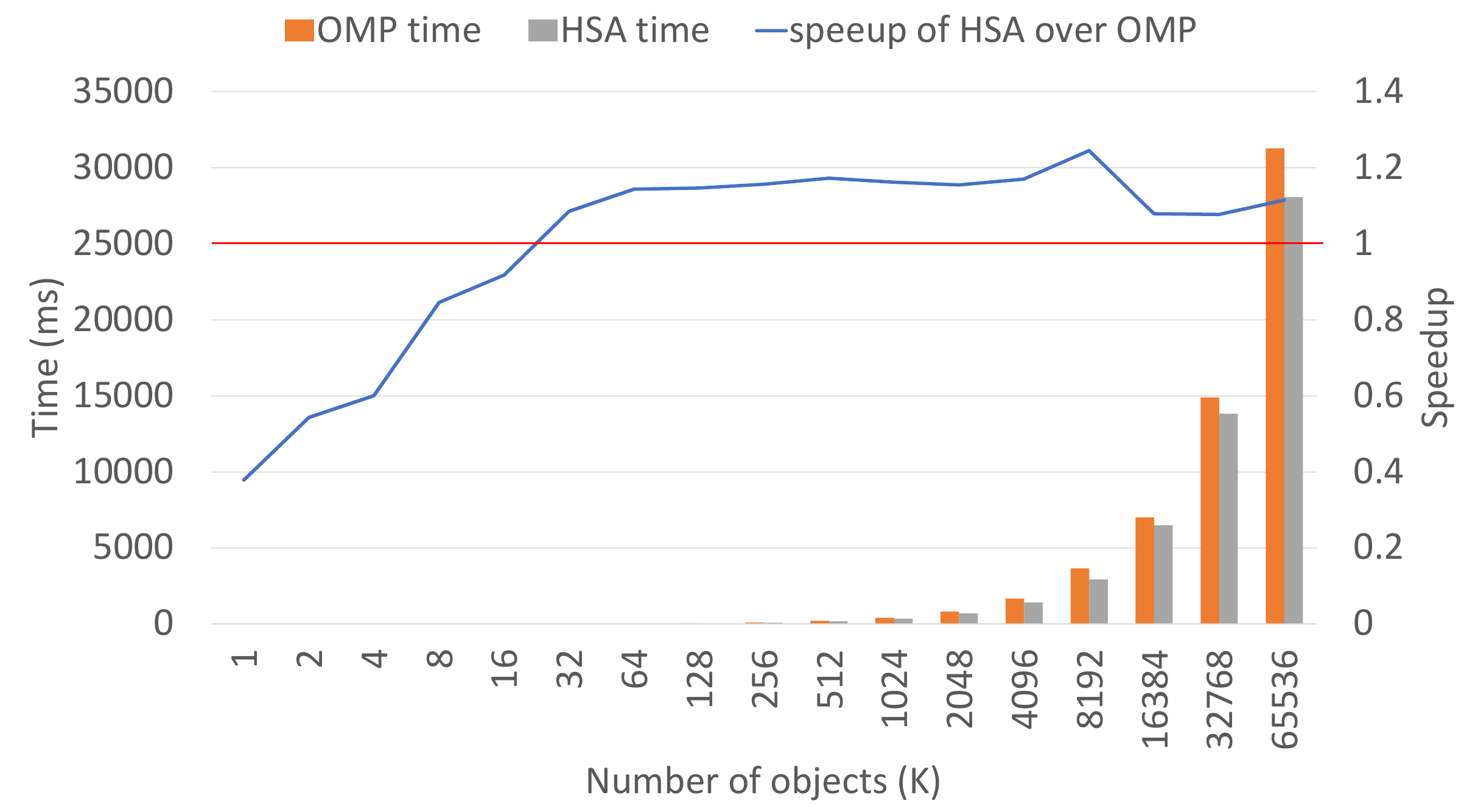} &   \includegraphics[scale=0.3]{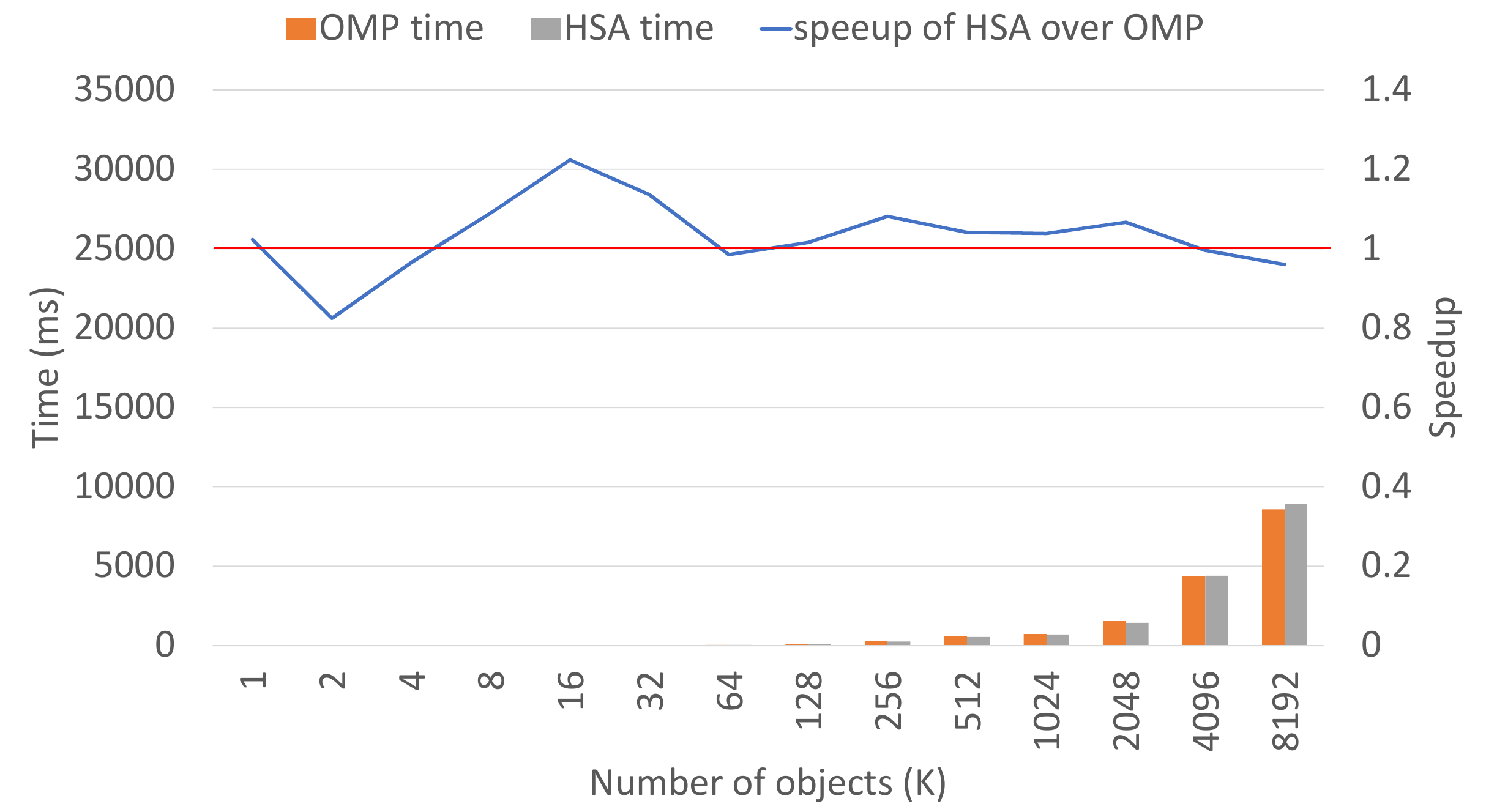} \\
(a) X = 8 & (b) X = 40 \\
 \includegraphics[scale=0.3]{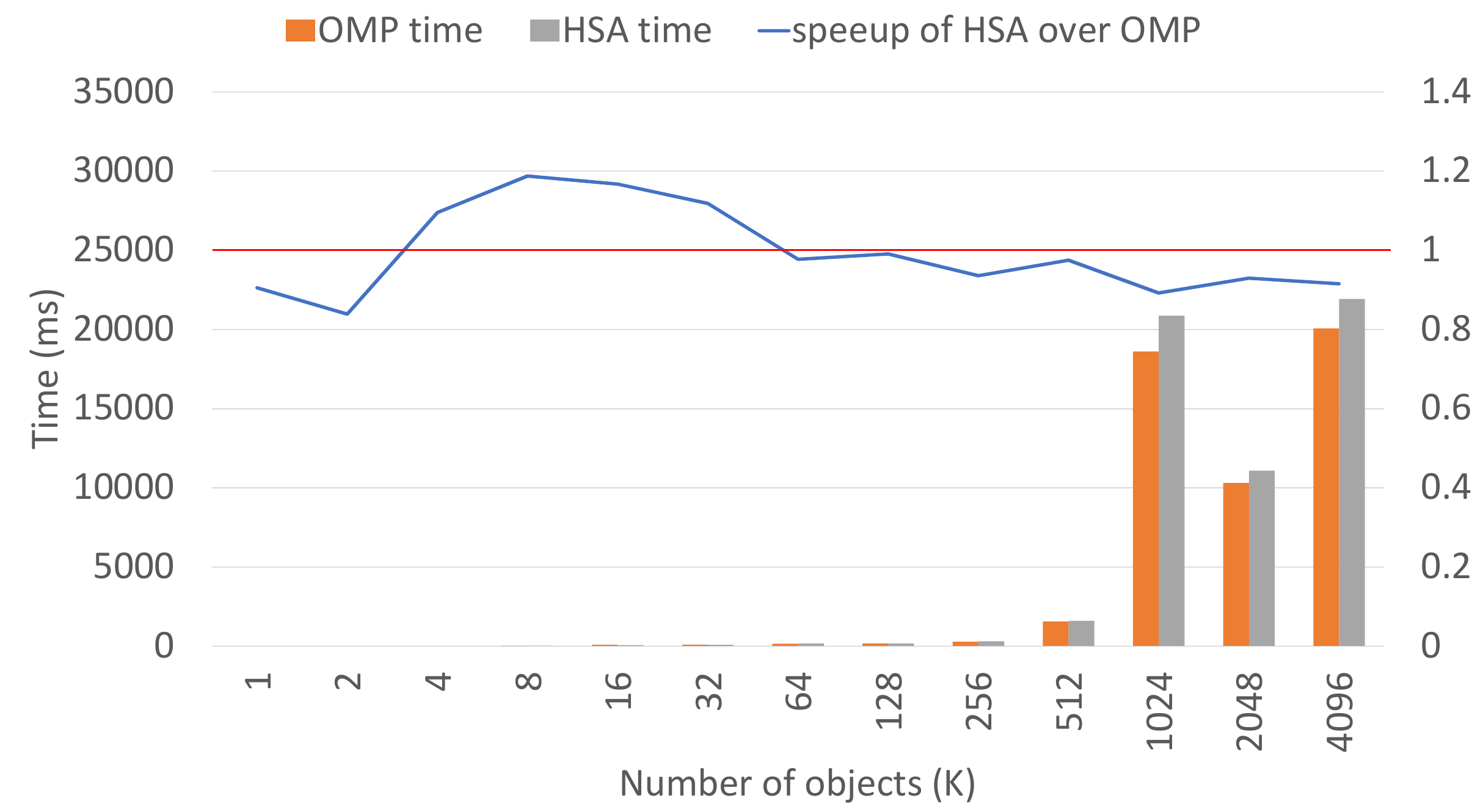} &   \includegraphics[scale=0.3]{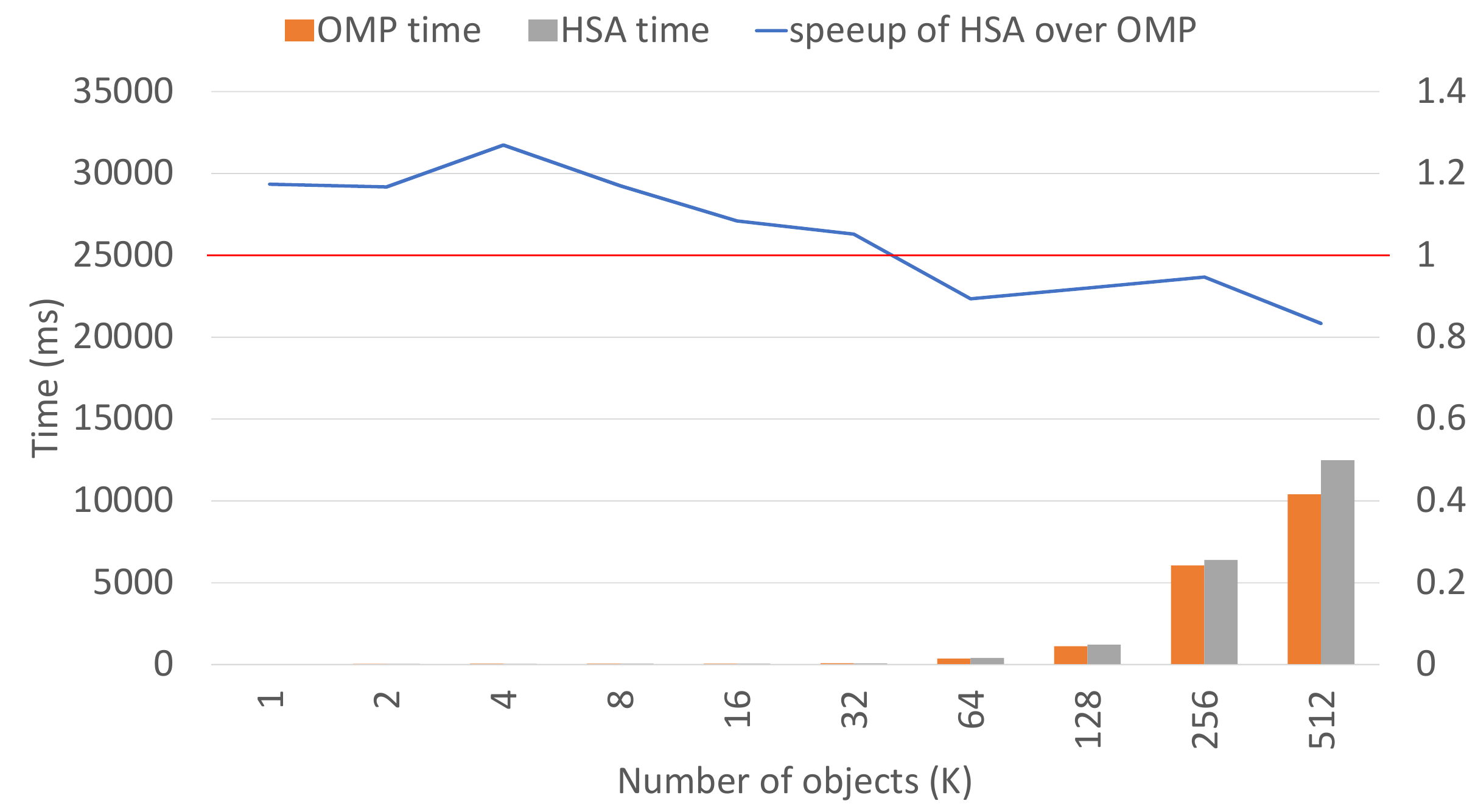} \\
(c) X = 200 & (d) X = 1000 \\
\end{tabular}
\caption{The time to traverse a heap object graph with increasing number of objects. A single atomic compare and swap instruction is added in each object access. The shape of the graph is randomly constructed with an average number edges = X. Hence, each object will have X links on average to other random objects.}
\label{fig:markingGPUatomic}
\end{figure*}

The previous results do not consider the case of concurrent marking. Concurrent marking requires atomic operations to ensure that mutator threads do not conflict with marking threads when they need to modify references simultaneously. Figure~\ref{fig:markingGPUatomic} shows the same experiments shown in figure~\ref{fig:markingGPU} but with atomically marking objects from the GPU. We use an atomic compare and exchange operation that is available for use in HSA and allows for a coherent CPU-GPU view and atomicity of the data that will be accessed. Concurrent marking using only CPU threads uses this atomic operation to update the mark bit when traversing objects. Hence, we apply the same technique for concurrent CPU-GPU marking. However, the cost of such operation is prohibitive. We can see from the figure that the use of atomic operations quickly diminishes the benefits that were observed by the results in figure~\ref{fig:markingGPU}. For these graphs, the performance from atomically marking objects on the GPU is 6.5x slower on average than non-atomic accesses; atomic marking on the CPU, however, is around 4.1x slower than non-atomic marking.

\subsection{\textbf{Programmability Challenges for Offloading Garbage Collection}}
\label{sec:prorammabilityGC}

One of our main concerns while investigating the potential of offloading garbage collection was programmability. In our previous work~\cite{dashti2017analyzing}, we concluded that HSA would provide the easiest (highest programmable) platform for seamless integration between CPU and GPU code. However, we ran into some compatibility issues. 

There are several challenges when it comes to trying to offload some of the GC work to the integrated GPU. Although we have a unified memory view from both CPU and GPU threads, there are still some software/programmability limitations. For example, GPU kernels are written in OpenCL. These kernels are compiled using a C compiler and do not know anything about the many classes and functions used in the JVM. 

It is not possible to simply pass objects of specific types to GPU kernels to work on them on the GPU. However, the GPU can access the same address space as the CPU and can access (read and write) any memory location that is passed to it.  For example, let's say that we decided to offload a specific function in the JVM GC code where this function manipulates some class objects. If the class is not a simple class (i.e, it contains virtual methods and uses inheritance), we cannot manipulate objects of this class on the GPU. In other words, we won't be able to perform parallel work on the GPU where we can modify or use these objects. So, we have to overlay the data structures that we need when offloading in a way where we can easily manipulate them when running on the GPU. This proved to be very tedious and error-prone.

Although with HSA "a pointer is a pointer" on both the CPU and the GPU, passing a pointer to a complex C++ class object does not allow us to seamlessly manipulate this object on the GPU. Although compiler toolchains such as HCC~\cite{HCC,amdrocm} allow the offloading of parallel work using HSA from within C++ code, it is not compatible with the JVM. We tried to compile the JVM using HCC (which is LLVM-based) but failed to produce a successfully built JVM. 

Therefore, it is essential to identify and be able to manipulate and represent all relevant data structures and functions for a specific task to be able to use them on the GPU. However, we have found that the engineering effort required to convert GC code to run on GPU is massive. Current GPU programming frameworks don't provide a clean and easy way of manipulating JVM C++ objects on the GPU side. Although other frameworks exist which might improve programmability such as SYCL~\cite{syclkhronos,hipsycl}, we are not aware of their compatibility with the JVM on our testing platform, so we leave their investigation for a future work.

\section{\textbf{Related Work}}

The idea of offloading garbage collection to specialized processors has been explored in a number of previous works. GPUs were first proposed to run garbage collection tasks in~\cite{jiva2012gpu}. The patent only provides a general description of coordinating garbage collection between a CPU and GPU, and suggests that the GPU is a potential candidate for offloading such a task since it is highly parallel. There are no design and implementation details to verify these claims. Furthermore, the authors in~\cite{veldema2011iterative} provide an implementation of a stop-the-world mark and sweep garbage collector for the GPU. They only provide a garbage collector for GPU tasks, and do not offload a garbage collector to the GPU. In other words, they provide an implicit (or managed) memory management for GPU memory allocations, which is different that what we propose.

Two related works tried to accelerate the marking phase of a stop-the-world serial garbage collection on integrated but old systems that did not have the utilities of unified memory~\cite{maas2012gpus,nasre2016fastcollect}. They had to overcome the shortcomings of these systems by creating a copy of the heap object graph that is kept consistent between the CPU and the GPU. Also, they only accelerated part of the marking phase of SerialGC and only considered Dacapo benchmarks. Our analysis and solution in this work is different in that we try to see if the programmability advantage of unified-memory integrated systems allows us to create a general framework for offloading any tasks in GC without any custom workarounds.

One further work explored the scheduling of garbage collection on heterogeneous multicore systems~\cite{akram2016boosting}. To our knowledge, this is the only work that explored distributing concurrent garbage collection tasks between two types of processors. However, their system is a single-ISA bigLittle architecture where the same type of CPU cores exist but a subset of the cores is in-order and low-power, and the other is out-of-order with higher performance. Therefore, the aim of their work is to choose where to run garbage collection based on metrics that measure the criticality of GC. Our work is different as we aim to offload parallel phases of garbage collection to an accelerator (GPU), which presents many programmability challenges as opposed to single-ISA systems.

 
GPUs have been used to accelerate general tasks that share some behaviors with garbage collation tasks. For example, graph traversals have been shown to provide good performance when accelerated on GPUs~\cite{merrill2012scalable, liu2015enterprise}. These works use elaborate techniques to demonstrate the usefulness of performing graph operations on GPU. We, however, explore unified integrated systems where one can simply utilize the GPU to perform basic graph operations such as BFS, which is the main component during the marking phases in GC.

There is a large body of works that provide general analysis of garbage collection such as~\cite{hertz2005quantifying, yu2016performance, blackburn2004myths, hussein2015impact}. In our work we provide experimentation and analysis relevant to our testing platform. Furthermore, there is a number of related works that target redesigning garbage collection algorithms to adapt them to specific types of applications such as big data~\cite{maas2015trash,maas2016taurus,nguyen2016yak,bruno2017ng2c}, or specific hardware such as NUMA systems~\cite{gidra2011assessing,gidra2013study,gidra2015numagic}. 


\section{\textbf{Conclusion}}
In this paper, we looked at a number of aspects of garbage collection and investigated the possibility of offloading such a subsystem from a programmability perspective on a heterogeneous platform. Programming languages and runtimes will have to utilize the heterogeneity of such platforms. Therefore, we presented a case study of such integration by implementing a module that is integrated into the JVM for general purpose offloading of parallel tasks from within the JVM. We used this module to offload some garbage collection tasks to study the viability of such offloading.

We conclude that non-optimized (having programmability in mind) GPU offloading of parallel garbage collection tasks does not yield any performance benefits on current integrated CPU-GPU systems. Although previous work has shown some promising results for running the marking phase of garbage collection on GPUs, they rely on custom solutions that are specifically tailored to certain systems and collectors. There is still a lot of limitations both in software and hardware which inhibit any benefits from trying to accelerate garbage collection tasks on integrated CPU-GPU systems. Although there might not be any performance advantage for GC, using the GPU as an additional processor might still be useful for other tasks.

\bibliographystyle{plain}
\bibliography{references}

\end{document}